\documentclass[a4paper,11pt,twoside]{article}
\usepackage[section]{algorithm}
\usepackage{algorithmic}

\usepackage{amsmath}
\usepackage{amsfonts}
\usepackage{amssymb}
\usepackage{graphicx,subfigure}
\usepackage{url}
\usepackage[standard,amsmath,thmmarks,thref,framed]{ntheorem}
\usepackage{dsfont} 
\usepackage{stmaryrd} 
\usepackage{alltt}
\date{}

\theoremheaderfont{\sc}\theorembodyfont{\upshape}
\theoremnumbering{arabic}
\theoremseparator{.}
\theoremsymbol{\ensuremath{\diamondsuit}}
\newtheorem{proof2}{proof}

\begin{document}

\newpage

\begin{center}
{\huge{Hash functions using chaotic iterations}}

\vspace{0.3cm}

{\bf Jacques M. Bahi, Christophe Guyeux}

jacques.bahi@univ-fcomte.fr \\
christophe.guyeux@univ-fcomte.fr \\
\end{center}

\vspace{0.5cm}

\noindent {\bf ABSTRACT} \\

In this paper, a novel formulation of discrete chaotic iterations in the field of dynamical systems is given. Their topological properties are studied: it is \linebreak mathematically proved that, under some conditions, these iterations have a chaotic behavior in the meaning of Devaney. This chaotic behavior allows us to propose a way to generate new hash functions.
An illustration example is detailed in order to show how to use our theoretical study in practice.
\\


\vspace{0.5cm}

\section{INTRODUCTION}

Chaotic iterations have been introduced on the one hand by Chazan and Miranker~\cite{Chazan69} in a numerical analysis context and on the other hand by Robert ~\cite{Robert1986} in the discrete dynamical systems framework. The goal was to derive sufficient conditions ensuring the convergence (or the stability) of such iterations. In this paper, a new point of view is presented: the goal here is to study the conditions under which these iterations admit a chaotic behavior. Contrary to the previous studies, convergence or stability are not sought.

This article presents the research results related to this question. %
We prove that under some conditions, discrete chaotic iterations
produce chaos, precisely, they produce topological chaos in the sense of
Devaney. This topological chaos is a rigorous and well studied
framework in the field of mathematical theory of chaos.

Behind the theoretical interest connecting the field of the chaotic discrete
iterations and the one of topological chaos, our study gives a framework
making it possible to create hash functions that can be mathematically
evaluated and compared.

\medskip

A hash function is a transformation that takes a variable-size input and returns a fixed-size string, which is called the hash value. They are used to speed up table lookup or data comparison tasks and for digital signature. Hash functions, such as MD5 or SHA-256, can be described by discrete
iterations on a finite set. In this paper, the elements of this finite set
are called cells. These cells represent the blocks of the text to which the
hash function will be applied.

Some required qualities for hash functions such as the avalanche effect, resistance to collisions and unpredictability
 can be mathematically described by notions
from the theory of topological chaos, namely, sensitivity, transitivity,
entropy and expansivity \cite{Li75}, \cite{Dev89}, \cite{Knudsen1994a}.
These concepts are approached but non deepened in
this article. More detailed studies will be carried out in forthcoming
articles.

\medskip

This study is the first of a series we intend to carry out. We think that
the mathematical framework in which we are placed offers interesting new
tools allowing the conception, the comparison and the evaluation of new
algorithms in computer security framework, not only hash functions.

\bigskip

The rest of the paper is organized as follows.\newline
The first next section is devoted to some recalls on two distinct domains: topological chaos and discrete chaotic iterations.%
\newline
The third and fourth sections constitute the theoretical study of the present paper. The topological framework is defined and the proof that chaotic iterations have a topological chaos behavior is given.

Section 5 details how it is possible to apply chaotic results in the computer science framework. Its following
section contains the application to hash functions and an illustration example. 
The paper ends by some discussions and future work.

\section{BASIC RECALLS}

This section is devoted to basic definitions and terminologies in the field
of topological chaos and in the one of chaotic iterations.

\subsection{Devaney's chaotic dynamical systems}

Consider a metric space $(\mathcal{X},d)$ and a continuous function $f:%
\mathcal{X}\longrightarrow \mathcal{X}$.

\begin{definition}
$f$ is said to be \emph{topologically transitive} if, for any pair of open
sets $U,V \subset \mathcal{X}$, there exists $k>0$ such that $f^k(U) \cap V
\neq \varnothing$.
\end{definition}

\begin{definition}
An element (a point) $x$ is a \emph{periodic element} (point) for $f$ of period $%
n\in \mathds{N},$ if $f^{n}(x)=x$. The set of periodic points of $f$ is denoted $%
Per(f).$
\end{definition}

\begin{definition}
$(\mathcal{X},f)$ is said to be \emph{regular} if the set of periodic points
is dense in $\mathcal{X}$,
\begin{equation*}
\forall x\in \mathcal{X},\forall \varepsilon >0,\exists p\in
Per(f),d(x,p)\leqslant \varepsilon .
\end{equation*}
\end{definition}

\begin{definition}
\label{sensitivity} $f$ has \emph{sensitive dependence on initial conditions}
if there exists $\delta >0$ such that, for any $x\in \mathcal{X}$ and any
neighborhood $V$ of $x$, there exists $y\in V$ and $n\geqslant 0$ such that $%
|f^{n}(x)-f^{n}(y)|>\delta $.

$\delta$ is called the \emph{constant of sensitivity} of $f$.
\end{definition}

Let us now recall the definition of a chaotic topological system, in the
sense of Devaney~\cite{Dev89}:

\begin{definition}
$f:\mathcal{X}\longrightarrow \mathcal{X}$ is said to be \emph{chaotic} on $%
\mathcal{X}$ if,

\begin{enumerate}
\item $f$ has sensitive dependence on initial conditions,

\item $f$ is topologically transitive,

\item $(\mathcal{X},f)$ is regular.
\end{enumerate}
\end{definition}

Therefore, quoting Robert Devaney: ``A chaotic map possesses three
ingredients: unpredictability, indecomposability and an element of
regularity. A chaotic system is unpredictable because of the sensitive
dependence on initial conditions. It cannot be broken down or decomposed
into two subsystems, because of topological transitivity. And, in the midst
of this random behavior, we nevertheless have an element of regularity,
namely the periodic points which are dense.'' Fundamentally different behaviors are then possible and occurs with an unpredictably way.

\subsection{Chaotic iterations}

In the sequel $S^{n}$ denotes the $n^{th}$ term of a sequence $S$, $V_{i}$
denotes the $i^{th}$ component of a vector $V$ and $f^{k}=f\circ ...\circ f$
denotes the $k^{th}$ composition of a function $f$. Finally, the following
notation is used: $\llbracket1;N\rrbracket=\{1,2,\hdots,N\}$.

Let us consider a \emph{system} of a finite number $\mathsf{N}$ of
\emph{cells}, so that each cell has a boolean \emph{state}. Then a sequence
of length $\mathsf{N}$ of boolean states of the cells corresponds to a
particular \emph{state of the system}. A sequence which elements belong in $\llbracket 1;\mathsf{N} \rrbracket $ is called a \emph{strategy}. The set of all
strategies is denoted by $\mathbb{S}.$

\begin{definition}
Let $S\in \mathbb{S}$. The \emph{shift} function is defined by $\sigma
:(S^{n})_{n\in \mathds{N}}\in \mathbb{S}\longrightarrow (S^{n+1})_{n\in %
\mathds{N}}\in \mathbb{S}$ and the \emph{initial function} $i$ is the map
which associates to a sequence, its first term: $i:(S^{n})_{n\in \mathds{N}%
}\in \mathbb{S}\longrightarrow S^{0}\in \llbracket1;\mathsf{N}\rrbracket$.
\end{definition}

\begin{definition}
The set $\mathds{B}$ denoting $\{0,1\}$, let $f:\mathds{B}^{\mathsf{N}%
}\longrightarrow \mathds{B}^{\mathsf{N}}$ be a function and $S\in \mathbb{S}
$ be a strategy. Then, the so-called \emph{chaotic iterations} are defined by

\begin{equation}
\left.
\begin{array}{l}
x^0\in \mathds{B}^{\mathsf{N}}, \\
\forall n\in \mathds{N}^{\ast },\forall i\in \llbracket1;\mathsf{N}\rrbracket%
,x_i^n=\left\{
\begin{array}{ll}
x_i^{n-1} & \text{ if }S^n\neq i \\
\left(f(x^n)\right)_{S^n} & \text{ if }S^n=i.%
\end{array}%
\right.%
\end{array}%
\right.  \label{chaotic iterations}
\end{equation}
\end{definition}

In other words, at the $n^{th}$ iteration, only the $S^{n}-$th cell is
\textquotedblleft iterated\textquotedblright . Note that in a more general
formulation, $S^n$ can be a subset of components and $f(x^{n})_{S^{n}}$ can
be replaced by $f(x^{k})_{S^{n}}$, where $k\leqslant n$, describing for
example delays transmission (see \emph{e.g.}~\cite{Bahi1999}, or \cite{Bahi2008bis}). For the
general definition of such chaotic iterations, see, e.g.~\cite{Robert1986}.

\section{THE NEW TOPOLOGICAL SPACE}

In this section we will put our study in a topological context by defining a
suitable metric space where chaotic iterations are continuous.

\subsection{Defining the iteration function and the phase space}

\label{Defining}

Denote by $\delta $ the \emph{discrete boolean metric}, $\delta
(x,y)=0\Leftrightarrow x=y.$ Given a function $f$, define the function

\begin{equation*}
\begin{array}{lrll}
F_{f}: & \llbracket1;\mathsf{N}\rrbracket\times \mathds{B}^{\mathsf{N}} &
\longrightarrow & \mathds{B}^{\mathsf{N}} \\
& (k,E) & \longmapsto & \left( E_{j}.\delta (k,j)+f(E)_{k}.\overline{\delta
(k,j)}\right) _{j\in \llbracket1;\mathsf{N}\rrbracket},%
\end{array}%
\end{equation*}%

\noindent where + and . are the boolean addition and product operations.

Consider the phase space:

\begin{equation*}
\mathcal{X} = \llbracket 1 ; \mathsf{N} \rrbracket^\mathds{N} \times
\mathds{B}^\mathsf{N},
\end{equation*}

\noindent and the map defined on $\mathcal{X}$:
\begin{equation}
G_f\left(S,E\right) = \left(\sigma(S), F_f(i(S),E)\right).  \label{Gf}
\end{equation}

\noindent Then the chaotic iterations defined in (\ref%
{chaotic iterations}) can be described by the following iterations
\begin{equation*}
\left\{
\begin{array}{l}
X^0 \in \mathcal{X} \\
X^{k+1}=G_{f}(X^k).%
\end{array}%
\right.
\end{equation*}%
\bigskip

\subsection{Cardinality of $\mathcal{X}$}

By comparing $\mathbb{S}$ and $\mathds{R}$, we have the result.

\begin{proposition}
The phase space $\mathcal{X}$ has, at least, the cardinality of the \linebreak continuum.
\end{proposition}

\begin{proof2}
Let $\varphi$ be the map which transforms a strategy into the binary representation of an element in $[0,1$[, as follows. If the $n^{th}$ term of the strategy is 0, then the $n^{th}$ associated digit is 0, else it is equal to 1.

With this construction, $\varphi : \llbracket 1 ; \mathsf{N} \rrbracket^\mathds{N} \longrightarrow  [0,1]$ is onto. But $]0,1[$ is isomorphic to $\mathds{R}$ ($x \in ]0,1[\mapsto tan(\pi(x-\frac{1}{2}))$ is an isomorphism), so the cardinal of $\llbracket 1 ; \mathsf{N} \rrbracket^\mathds{N}$ is greater or equal than the cardinal of $\mathds{R}$. So the cardinal of the Cartesian product $\mathcal{X} = \llbracket 1 ; \mathsf{N} \rrbracket^\mathds{N} \times \mathds{B}^\mathsf{N}$ is greater or equal to the cardinal of $\mathds{R}$.
\end{proof2}

\begin{remark}
This result is independent on the number of cells of the system.
\end{remark}

\subsection{A new distance}

We define a new distance between two points $X = (S,E), Y = (\check{S},\check{E})\in
\mathcal{X}$ by%
\begin{equation*}
d(X,Y)=d_{e}(E,\check{E})+d_{s}(S,\check{S}),
\end{equation*}

\noindent where

\begin{equation*}
\left\{
\begin{array}{lll}
\displaystyle{d_{e}(E,\check{E})} & = & \displaystyle{\sum_{k=1}^{\mathsf{N}%
}\delta (E_{k},\check{E}_{k})}, \\
\displaystyle{d_{s}(S,\check{S})} & = & \displaystyle{\dfrac{9}{\mathsf{N}}%
\sum_{k=1}^{\infty }\dfrac{|S^k-\check{S}^k|}{10^{k}}}.%
\end{array}%
\right.
\end{equation*}

If the floor value $\lfloor d(X,Y)\rfloor $ is equal to $n$,
then the systems $E, \check{E}$ differ in $n$ cells. In addition, $d(X,Y) - \lfloor d(X,Y) \rfloor $ is a measure of the differences between strategies $S$ and $\check{S}$. More precisely, this floating part is less than $10^{-k}$ if and only if the first $k$
terms of the two strategies are equal. Moreover, if the $k^{th}$ digit is nonzero, then the $k^{th}$ terms of the two
strategies are different.

\subsection{Continuity of the iteration function}

To prove that chaotic iterations are an example of topological chaos in the
sense of Devaney ~\cite{Dev89}, $G_{f}$ must be continuous on the metric
space $(\mathcal{X},d)$.

\begin{theorem}
$G_f$ is a continuous function.
\end{theorem}

\begin{proof2}
We use the sequential continuity.

Let $(S^n,E^n)_{n\in \mathds{N}}$ be a sequence of the phase space $%
\mathcal{X}$, which converges to $(S,E)$. We will prove that $\left(
G_{f}(S^n,E^n)\right) _{n\in \mathds{N}}$ converges to $\left(
G_{f}(S,E)\right) $. Let us recall that for all $n$, $S^n$ is a strategy,
thus, we consider a sequence of strategies (\emph{i.e.} a sequence of
sequences).\newline
As $d((S^n,E^n);(S,E))$ converges to 0, each distance $d_{e}(E^n,E)$ and $d_{s}(S^n,S)$ converges
to 0. But $d_{e}(E^n,E)$ is an integer, so $\exists n_{0}\in \mathds{N},$ $%
d_{e}(E^n,E)=0$ for any $n\geqslant n_{0}$.\newline
In other words, there exists a threshold $n_{0}\in \mathds{N}$ after which no
cell will change its state:
\begin{equation*}
\exists n_{0}\in \mathds{N},n\geqslant n_{0}\Longrightarrow E^n = E.
\end{equation*}%
In addition, $d_{s}(S^n,S)\longrightarrow 0,$ so $\exists n_{1}\in %
\mathds{N},d_{s}(S^n,S)<10^{-1}$ for all indexes greater than or equal to $%
n_{1}$. This means that for $n\geqslant n_{1}$, all the $S^n$ have the same
first term, which is $S^0$:%
\begin{equation*}
\forall n\geqslant n_{1},S_0^n=S_0.
\end{equation*}%
Thus, after the $max(n_{0},n_{1})^{th}$ term, states of $E^n$ and $E$ are
identical, and strategies $S^n$ and $S$ start with the same first term.\newline
Consequently, states of $G_{f}(S^n,E^n)$ and $G_{f}(S,E)$ are equal,
so, after the \linebreak $max(n_0, n_1)^{th}$ term, the distance $d$ between these two points is strictly less than 1.\newline
\noindent We now prove that the distance between $\left(
G_{f}(S^n,E^n)\right) $ and $\left( G_{f}(S,E)\right) $ is convergent to
0. Let $\varepsilon >0$. \medskip

\begin{itemize}
\item If $\varepsilon \geqslant 1$, we have seen that distance
between $\left( G_{f}(S^n,E^n)\right) $ and $\left( G_{f}(S,E)\right) $ is
strictly less than 1 after the $max(n_{0},n_{1})^{th}$ term (same state).
\medskip

\item If $\varepsilon <1$, then $\exists k\in \mathds{N},10^{-k}\geqslant
\varepsilon \geqslant 10^{-(k+1)}$. But $d_{s}(S^n,S)$ converges to 0, so
\begin{equation*}
\exists n_{2}\in \mathds{N},\forall n\geqslant
n_{2},d_{s}(S^n,S)<10^{-(k+2)},
\end{equation*}%
thus after $n_{2}$, the $k+2$ first terms of $S^n$ and $S$ are equal.
\end{itemize}

\noindent As a consequence, the $k+1$ first entries of the strategies of $%
G_{f}(S^n,E^n)$ and $G_{f}(S,E)$ are the same ($G_{f}$ is a shift of strategies) and due to the definition of $d_{s}$, the floating part of
the distance between $(S^n,E^n)$ and $(S,E)$ is strictly less than $%
10^{-(k+1)}\leqslant \varepsilon $.\bigskip \newline
In conclusion,
\begin{equation*}
\forall \varepsilon >0,\exists N_{0}=max(n_{0},n_{1},n_{2})\in \mathds{N}%
,\forall n\geqslant N_{0},d\left( G_{f}(S^n,E^n);G_{f}(S,E)\right)
\leqslant \varepsilon .
\end{equation*}
$G_{f}$ is consequently continuous.
\end{proof2}

In this section, we proved that chaotic iterations can be modelized as a
dynamical system in a topological space. In the next section, we show that
chaotic iterations are a case of topological chaos, in the sense of Devaney.

\section{DISCRETE CHAOTIC ITERATIONS AS TOPOLOGICAL CHAOS}

To prove that we are in the framework of Devaney's topological chaos, we
have to check the regularity, transitivity and sensitivity conditions.
We will prove that the vectorial logical negation function
\begin{equation}
f_{0}(x_{1},%
\hdots,x_{\mathsf{N}})=(\overline{x_{1}},\hdots,\overline{x_{\mathsf{N}}})
\label{f0}
\end{equation}

\noindent satisfies these hypotheses.

\subsection{Regularity}

\label{regularite}

\begin{proposition}
Periodic points of $G_{f_0}$ are dense in $\mathcal{X}$.
\end{proposition}

\begin{proof2}
Let $(\check{S}, \check{E})\in \mathcal{X}$ and $\varepsilon >0$. We are
looking for a periodic point $(\widetilde{S},\widetilde{E})$ satisfying $d((%
\check{S}, \check{E});(\widetilde{S},\widetilde{E}))<\varepsilon$.

As $\varepsilon$ can be strictly lesser than 1, we must choose $%
\widetilde{E} = \check{E}$. Let us define $k_0(\varepsilon) =\lfloor
log_{10}(\varepsilon )\rfloor +1$ and consider the set
\[
\mathcal{S}_{\check{S}, k_0(\varepsilon)} = \left\{ S \in \mathbb{S} / S^k =
\check{S}^k, \forall k \leqslant k_0(\varepsilon) \right\}.
\]

Then, $\forall S \in \mathcal{S}_{\check{S}, k_0(\varepsilon)}, d((S, \check{%
E});(\check{S}, \check{E})) < \varepsilon$. It remains to choose $\widetilde{%
S} \in \mathcal{S}_{\check{S}, k_0(\varepsilon)}$ such that $(\widetilde{S},%
\widetilde{E}) = (\widetilde{S},\check{E})$ is a periodic point for $%
G_{f_0}$.

Let $\mathcal{J} = \left\{ i \in \{1, 2, ..., \mathsf{N}\} / E_i \neq \check{%
E}_i, \text{ where } (S, E) = G_{f_0}^{k_0} (\check{S}, \check{E}) \right\}$%
, $i_0 = card(\mathcal{J})$ and $j_1 <j_2 < ... < j_{i_0}$ the elements of $%
\mathcal{J}$. Then, $\widetilde{S} \in \mathcal{S}_{\check{S},
k_0(\varepsilon)}$ defined by

\begin{itemize}
\item $\widetilde{S}^k = \check{S}^k$, if $k \leqslant k_0(\varepsilon)$,
\item $\widetilde{S}^k = j_{k-k_0(\varepsilon)}$, if $k \in
\{k_0(\varepsilon)+1, k_0(\varepsilon)+2, ..., k_0(\varepsilon)+i_0\}$,
\item and $\widetilde{S}^{k}=\widetilde{S}^{j}$, where $j\leqslant
k_{0}(\varepsilon )+i_{0}$ is satisfying $j\equiv k~(\text{mod }%
k_{0}(\varepsilon )+i_{0})$, if $k>k_{0}(\varepsilon )+i_{0}$,
\end{itemize}

\noindent is such that $%
(\widetilde{S},\widetilde{E})$ is a periodic point, of period $%
k_{0}(\varepsilon )+i_{0}$, which is $\varepsilon -$closed to $(\check{S},%
\check{E})$.\newline As a conclusion, $(\mathcal{X},G_{f_{0}})$ is
regular.
\end{proof2}

\subsection{Transitivity}

\begin{proposition}
$(\mathcal{X},G_{f_0})$ is topologically transitive.
\end{proposition}

\begin{proof2}
Let us define $\mathcal{E}:\mathcal{X}\rightarrow \mathbb{B}^{\mathsf{N}},$
such that $\mathcal{E(}S,E)=E.$ Let \linebreak $\mathcal{B}_{A}=\mathcal{B}(X_{A},r_{A})
$ and $\mathcal{B}_{B}=\mathcal{B}(X_{B},r_{B})$ be two open balls of $%
\mathcal{X}$, with\linebreak $X_{A}=(S_{A},E_{A})$ and $X_{B}=(S_{B},E_{B})$. We are
looking for $\widetilde{X}=(\widetilde{S},\widetilde{E})$ in $\mathcal{B}_{A}
$ such that $\exists n_{0}\in \mathbb{N},G_{f_{0}}^{n_{0}}(\widetilde{X})\in
\mathcal{B}_{B}$.\newline
$\widetilde{X}$ must be in $\mathcal{B}_{A}$ and $r_{A}$ can be strictly
lesser than 1, so $\widetilde{E}=E_{A}$. Let $k_{0}=\lfloor \log
_{10}(r_{A})+1\rfloor $. Then $\forall S\in \mathbb{S}$, if $%
S^{k}=S_{A}^{k},\forall k\leqslant k_{0}$, then $(S,\widetilde{E})\in
\mathcal{B}_{A}$. Let us notice $(\check{S},\check{E}%
)=G_{f_{0}}^{k_{0}}(S_{A},E_{A})$ and $c_{1},...,c_{k_{1}}$ the elements of
the set $\{i\in \llbracket1,\mathsf{N}\rrbracket/\check{E}_{i}\neq \mathcal{E%
}(X_{B})_{i}\}.$ So any point $X$ of the set
\[
\{(S,E_{A})\in \mathcal{X}/\forall k\leqslant k_{0},S^{k}=S_{A}^{k}\text{
and }\forall k\in \llbracket1,k_{1}\rrbracket,S^{k_{0}+k}=c_{k}\}
\]%
is satisfying $X\in \mathcal{B}_{A}$ and $\mathcal{E}\left(
G_{f_{0}}^{k_{0}+k_{1}}(X)\right) =E_{B}$.

\noindent Last, let us define $k_2 = \lfloor \log_{10}(r_B)\rfloor +1$. Then $%
\widetilde{X} = (\widetilde{S}, \widetilde{E}) \in \mathcal{X}$ defined by:

\begin{enumerate}
\item $\widetilde{X} = E_A$,
\item $\forall k \leqslant k_0, \widetilde{S}^k = S_A^k$,
\item  $\forall k \in \llbracket 1, k_1 \rrbracket,$ $\widetilde{S}^{k_0+k} =
c_k$,
\item $\forall k \in \mathbb{N}^*, \widetilde{S}^{k_0+k_1+k} = S_B^k$,
\end{enumerate}

\noindent is such that $\widetilde{X} \in \mathcal{B}_A$ and $G_{f_0}^{k_0+k_1}(%
\widetilde{X}) \in \mathcal{B}_B$.
\end{proof2}

\subsection{Sensitive dependence on initial conditions}

\begin{proposition}
$(\mathcal{X},G_{f_0})$ has sensitive dependence on initial conditions.
\end{proposition}

\begin{proof2}
Banks \emph{et al.} proved in ~\cite{Banks92} that having sensitive dependence is a consequence of being regular and topologically transitive.
\end{proof2}





\subsection{Devaney's Chaos}

In conclusion, $(\mathcal{X},G_{f_0})$ is topologically transitive, regular
and has sensitive dependence on initial conditions. Then we have the result.

\begin{theorem}
$G_{f_0}$ is a chaotic map on $(\mathcal{X},d)$ in the sense of Devaney.
\end{theorem}
%

\begin{remark}
We have proved that the set of the iterate functions $f$ such that
$(\mathcal{X}, G_f)$ is chaotic (in the meaning of Devaney), is a nonempty set.
In a future work, we will give a characterization of this set.
\end{remark}

\section{CHAOS IN COMPUTER SCIENCE}

\label{Concerning}

It is worthwhile to notice that even if the set of machine numbers is
finite, we deal with the \emph{infinite} set of strategies that have a finite but unbounded lengths.
Indeed, it is not necessary to store all the terms of the strategy in the
memory, only the $n^{th}$ term (an integer less than or equal to $\mathsf{N}$) of the
strategy has to be stored at the $n^{th}$ step, as it is illustrated in the
following example. Let us suppose that a given text is input from the
outside world into the computer character by character and that the current
term of the strategy is computed from the ASCII code of the current stored
character. Then, as the set of all possible texts of the outside world is
infinite and the number of their characters is unbounded, we have to deal
with an infinite set of finite but unbounded strategies.

Of course, the previous example is a simplistic illustrating example. A chaotic procedure should to be introduced to generate the terms of the strategy from the stream of characters.\newline

Then in the computer science framework, we also have to deal with a finite set of states of the form $\mathds{B}^\mathsf{N}$ and as stated before an infinite set $\mathbb{S}$ of strategies. The sole difference with the previous study is that, instead of being infinite the sequences of $S$ are finite with unbounded length.\newline
The proofs of continuity and transitivity are independent of the finiteness of the length of strategies (sequences of $%
\mathbb{S}$). 
In addition, it is possible to prove the sensitivity property in this situation. So even in the case of finite machine numbers, we have the two
fundamental properties of chaos: sensitivity and transitivity, which respectively implies unpredictability and indecomposability (see~\cite{Dev89}, p.50). The regularity supposes that the sequences are of infinite lengths. To obtain the analogous of regularity in the context of finite sets, we define below the notion of \emph{periodic but finite} sequences.
 \medskip

\begin{definition}
A strategy $S\in\mathbb{S}$ is said to be \emph{periodic but finite}
if $S$ is a finite sequence of length $n$ and if there exists a divisor $p$ of $n$, $p \neq n$,
 such that $\forall i \leqslant n-p, S^i = S^{i+p}$.
A point $(E,S) \in \mathcal{X}$ is said to be \emph{periodic but finite}, if its
strategy $S$ is periodic but finite.
\end{definition}

For example, $(1,2,1,2,1,2,1,2)$ ($p$=2) and $(2,2,2)$ ($p$=1), are periodic but finite. This
definition can be interpreted as the analogous of periodicity on finite strategies.
 Then, following the proof of regularity (section \ref{regularite}), it can
be proved that the set of periodic but finite points is dense on $\mathcal{X}$,
 hence obtaining a desired element of regularity in finite sets, as quoted by
 Devaney (\cite{Dev89}, p.50): two points arbitrary close to each other could have
completely different behaviors, the one could have a cyclic behavior as long as the
system iterates while the trajectory of the second could "visit" the whole phase space. It should be recalled that the regularity was introduced by Devaney in order to counteract the effects of sensitivity and transitivity: two points close to each other can have fundamental different behaviors.

\bigskip

In conclusion, even in the computer science framework our previous theory
applies. In what follows, an example of the use of chaotic iterations in
the field of computer science is given.

\section{HASH FUNCTIONS BASED ON TOPOLOGICAL CHAOS}

\subsection{Introduction}

The use of chaotic map to generate hash algorithm is a recent idea.
In \cite{Fei2005} for example, a digital signature algorithm  based on elliptic curve and chaotic mappings is
proposed to strengthen the security of an elliptic curve digital signature algorithm. Other examples of the generation of an hash function using chaotic maps can be found in \cite{Wang2003} and \cite{Peng2005}.

\bigskip

We define in this section a new way to construct hash functions based on chaotic iterations. As a consequence of the previous theory, generated hash functions satisfy the topological chaos property. Thus, this approach guarantees to obtain various desired properties in the domain of hash functions. For example, the avalanche criterion is closely linked to the sensitivity property.

\bigskip

The hash value will be the last state of some chaotic iterations: initial state $X_0$, finite strategy $S$ and iterate function must then be defined.

\subsection{Initial state}

The initial condition $X_0=\left( S,E\right) $ is composed by:

\begin{itemize}
\item A $\mathsf{N} = 256$ bits sequence $E$ obtained from the original text.

\item A chaotic strategy $S$.
\end{itemize}

\noindent In the sequel, we describe in detail how to obtain this initial condition.

\subsubsection{How to obtain $E$}

The first step of our algorithm is to transform the message in a normalized
256 bits sequence $E$. To illustrate this step, we take an example, our
original text is: \emph{The original text}

Each character of this string is replaced by its ASCII code (on 7 bits).
Then, we add a 1 to this string.

\bigskip

\begin{center}
\begin{alltt}
\noindent 10101001 10100011 00101010 00001101 11111100 10110100
\noindent 11100111 11010011 10111011 00001110 11000100 00011101
\noindent 00110010 11111000 11101001
\end{alltt}
\end{center}

\bigskip

So, the binary value (1111000) of the length of this string (120) is added, with another 1:

\bigskip

\begin{center}
\begin{alltt}
\noindent 10101001 10100011 00101010 00001101 11111100 10110100
\noindent 11100111 11010011 10111011 00001110 11000100 00011101
\noindent 00110010 11111000 11101001 11110001
\end{alltt}
\end{center}

\bigskip

The whole string is copied, but in the opposite direction. This gives:

\bigskip

\begin{center}
\begin{alltt}
\noindent 10101001 10100011 00101010 00001101 11111100 10110100
\noindent 11100111 11010011 10111011 00001110 11000100 00011101
\noindent 00110010 11111000 11101001 11110001 00011111 00101110
\noindent 00111110 10011001 01110000 01000110 11100001 10111011
\noindent 10010111 11001110 01011010 01111111 01100000 10101001
\noindent 10001011 0010101
\end{alltt}
\end{center}

\medskip So, we obtain a multiple of 512, by duplicating enough this string and truncating at the next multiple of 512.
This string, in which the whole original text is contained, is denoted by $D$.

\bigskip

Finally, we split our obtained string into blocks of 256 bits and apply to them the exclusive-or function, obtaining a 256 bits sequence.

\bigskip
\begin{alltt}
\noindent 11111010 11100101 01111110 00010110 00000101 11011101
\noindent 00101000 01110100 11001101 00010011 01001100 00100111
\noindent 01010111 00001001 00111010 00010011 00100001 01110010
\noindent 01000011 10101011 10010000 11001011 00100010 11001100
\noindent 10111000 01010010 11101110 10000001 10100001 11111010
\noindent 10011101 01111101
\end{alltt}

So, in the context of subsection (1), $\mathsf{N}=256$ and $E$ is the above obtained sequence of 256 bits.

\medskip

We now have the definitive length of our digest. Note that a lot of texts
have the same string. This is not a problem because the strategy we will
build will depends on the whole text.

Let us build now the strategy $S$.

\subsubsection{How to choose $S$}

To obtain the strategy $S$, an intermediate sequence $(u^n)$ is constructed from $D$, as follows:

\begin{enumerate}
\item $D$ is split into blocks of 8 bits. Then $u^n$ is the decimal value of the $n^{th}$ block.

\item A circular rotation of one bit to the left is applied to $D$ (the first bit of $D$ is put on the end of $D$). Then the new string is split into blocks of 8 bits another time. The decimal values of those blocks are added to $(u^n)$.

\item This operation is repeated again 6 times.
\end{enumerate}

\bigskip

It is now possible to build the strategy $S$:




\begin{equation*}
S^0 = u^0,~~~
S^n=(u^n+2\times S^{n-1}+n) ~(mod ~256).
\end{equation*}%

\noindent $S$ will be highly dependent to the changes of the original text, because \linebreak $\theta \longmapsto 2\theta ~(mod ~1)$ is known to be chaotic in the sense of Devaney \cite{Dev89}.

\subsubsection{How to construct the digest}

To construct the digest, chaotic iterations are done with initial state $X^0$,

\begin{equation*}
\begin{array}{rccc}
f: & \llbracket1,256\rrbracket & \longrightarrow  & \llbracket1,256\rrbracket
\\
& (E_1,\hdots,E_{256}) & \longmapsto  & (\overline{E_1},\hdots,\overline{%
E_{256}}),%
\end{array}%
\end{equation*}%
\noindent as iterate function and $S$ for the chaotic strategy.

\bigskip

\noindent The result of those iterations is a 256 bits vector. Its components are taken 4 per 4 bits and translated into hexadecimal numbers, to obtain the hash value:

\medskip
\begin{alltt}
\noindent 63A88CB6AF0B18E3BE828F9BDA4596A6A13DFE38440AB9557DA1C0C6B1EDBDBD
\end{alltt}

\bigskip

As a comparison if instead of considering the text \textquotedblleft \textit{%
The original text}\textquotedblright\ we took \textquotedblleft \textit{the original text}\textquotedblright , the hash function returns:

\medskip
\begin{alltt}
\noindent 33E0DFB5BB1D88C924D2AF80B14FF5A7B1A3DEF9D0E831194BD814C8A3B948B3
\end{alltt}

\bigskip

In this paper, the generation of hash value is done with the vectorial boolean negation $f_{0} $ defined in eq. (\ref{f0}). Nevertheless, the procedure remains general and can be applied with any function $f$ such that $G_f$ is chaotic.

\medskip

In the following subsection, a complete example of the proceeding is given.

\subsection{Application example}


Consider the following message \cite{Poe}:
\begin{center}
\begin{alltt}
                 Wanderers in that happy valley,
                       Through two luminous windows, saw
                 Spirits moving musically,
                       To a lute's well-tuned law,
                 Round about a throne where, sitting
                       (Porphyrogene !)
                 In state his glory well befitting,
                       The ruler of the realm was seen.

                 And all with pearl and ruby glowing
                       Was the fair palace door,
                 Through which came flowing, flowing,
                       And sparkling evermore,
                 A troop of Echoes, whose sweet duty
                       Was but to sing,
                 In voices of surpassing beauty,
                       The wit and wisdom of their king.
\end{alltt}
\end{center}

\bigskip

Its hash value is:
\begin{alltt}
\noindent FF51DA4E7E50FBA7A8DC6858E9EC3353BDE2E465E1A6A1B03BEAA12A4AD694FB
\end{alltt}
\bigskip

As a comparison, if an additional space is put before ``       Was the fair palace door,'' the hash value will be:
\begin{alltt}
\noindent 03ABFA49B834D529669CFC1AEEC13E14EA5FFD2349582380BCBDBF8400017445
\end{alltt}
\bigskip

\noindent and if "Echoes" is replaced by "echoes" in the original text:

\begin{alltt}
\noindent FE54777C52D373B7AED2EA5ACAD422B5B563BB3B91E8FCB48AAE9331DAC54A9B
\end{alltt}

\medskip

Those examples give an illustration of the avalanche effect obtained by this algorithm. A more complete study of the properties possessed by our hash functions and
resistance under collisions will be studied in a future work.

\section{CONCLUSION}

We proved that discrete chaotic iterations behave as Devaney's topological chaos if the iteration function is the vectorial boolean negation function.\newline
We applied these results to the generation of new hash functions. The vectorial boolean negation function has been chosen here, but the process remains general and other iterate functions $f$ can be used. The sole condition is to prove that $G_f$ satisfies Devaney's chaos property.\newline
By considering hash functions as an application of our theory, we shown
how some desirable aspects in computer security field such as unpredictability,
sensitivity to initial conditions, mixture and disorder can be
mathematically guaranteed and even quantified by mathematical tools.\newline

Theory of chaos recalls us that simple functions can have, when iterated, a
very complex behavior, while some complicated functions could have
foreseeable iterations. This is why it is important to have tools for
evaluating desired properties.\newline
Our simple function may be replaced by other "chaotic" functions which
can be evaluated with quantitative tools, like the constant of sensitivity. Another
important parameter is the choice of the strategy $S$. We proposed a
particular strategy that can be easily improved by multiple ways.%
\newline

Much work remains to be made. For example we are convinced that the good
comprehension of the transitivity property, enables to study the problem of
collisions in hash functions. \newline
In future work we plan to investigate other forms of chaos such as Li-York \cite{Li75} or Knudsen \cite{Knudsen1994a}
chaos. Other quantitative and qualitative tools
like expansivity or entropy (see e.g. \cite{Bowen}) will be explored and the domain of
applications of our theoretical concepts will be enlarged.




\nocite{*}

\bibliography{mabase}
\bibliographystyle{unsrt}

\end{document}